\documentclass[aps,floatfix,superscriptaddress,citeautoscript,twocolumn]{revtex4-2}
\usepackage{graphicx} 
\usepackage{xcolor}
\usepackage{comment}
\usepackage{hyperref}
\hypersetup{colorlinks=true, pdfstartview=FitV, linkcolor=blue, citecolor=black, plainpages=false, pdfpagelabels=true, urlcolor=blue}
\usepackage[all]{hypcap}  
\usepackage{amsmath}

\newcommand{\pgithirteen}[0]{\affiliation{Institute for Functional Quantum System (PGI-13), Forschungszentrum J\"ulich, 52425 J\"ulich, Germany}}
\newcommand{\rwth}[0]{\affiliation{Department of Physics, RWTH Aachen University, 52074 Aachen, Germany}}

\begin{document}
\title{Flux Crosstalk Calibration from Zero Knowledge to High Precision in Fully Tunable Superconducting Circuits}

\author{Y. Gao}
\email[]{y.gao@fz-juelich.de}
\pgithirteen
\rwth
\author{A. Galicia}
\pgithirteen
\rwth
\author{J. R. Guimar\~aes}
\pgithirteen
\rwth
\author{Y.~Haddad}
\pgithirteen
\rwth
\author{Y. Liu}
\pgithirteen
\author{J. Cereijo}
\pgithirteen
\rwth
\author{D.~A.~Volkov}
\pgithirteen
\rwth
\author{H.~Bhardwaj}
\pgithirteen
\rwth
\author{M. Neis}
\pgithirteen
\rwth
\author{M. Jerger}
\pgithirteen
\author{P. A. Bushev}
\pgithirteen
\author{R. Barends}
\email[]{r.barends@fz-juelich.de}
\pgithirteen
\rwth

\begin{abstract}
Precise and independent flux control is essential for scalable superconducting processors built from frequency-tunable qubits and couplers, but flux crosstalk undermines it. Conventional calibration methods rely either on dedicated readout channels, which tunable couplers typically lack, or on readout schemes requiring extensive prior spectroscopy of the qubit and coupler. We demonstrate two independent methods that characterize coupler flux crosstalk without directly driving or measuring the couplers. The first method is zero-knowledge characterization, which exploits intrinsic coupler–resonator coupling to yield rapid and accurate estimates with no prior qubit or coupler spectroscopy, accelerating large scale bring-up. The second method is symmetric branch cancellation, which isolates crosstalk from other dispersive effects allowing for quantifying with high precision. Combined with an improved iterative procedure, these methods suppress residual coupler crosstalk below $5\times10^{-5}$, providing a complete route from zero-knowledge initialization to high-precision crosstalk characterization in fully tunable architectures.

\end{abstract}
\maketitle

Scalable quantum processors increasingly demand precise and independent control for practical applications~\cite{preskill_review,roadmap}. Superconducting circuits are a particularly versatile platform for scalable processors where frequency tunable elements are the key ingredients, including transmon~\cite{Koch_transmon,xmon}, fluxonium~\cite{fluxonium}, and other unconventional qubits~\cite{zero-pi}. Individual tunability of qubits and couplers allows qubit frequencies to be shifted away from detrimental defects while providing flexible control over the coupling between qubits~\cite{TLS,tunable_coupler}. 
To achieve such independent addressability, flux control lines are commonly employed individually, and have demonstrated high fidelity operations. 
However, the complex interplay among different control lines leads to inadvertent flux crosstalk.
Such crosstalk compromises the independence of the control, deteriorates the qubit performance, and degrades the simultaneous gate fidelities. This issue becomes even more prominent with increasing control density and integration level.

Various chip design and fabrication techniques have been developed to suppress flux crosstalk at the source, including flip-chip architectures, airbridge interconnects, and differential fluxline configurations~\cite{float_fluxline,flipchip,airbridge,air_bridge}. Nevertheless, reducing crosstalk to a negligible level through layout optimization alone remains impractical. Consequently, in-situ characterization of the crosstalk matrix followed by an active compensation is generally indispensable. 
The flux crosstalk typically induces unintended flux offsets in affected circuit elements, manifesting as shifts in their measured frequency. Previous works therefore rely on the direct spectroscopy of the affected elements to characterize the flux crosstalk~\cite{fixed_coupling,coupler_resonator,coupler_resonator2,Symmetric}. However, these approaches are not directly applicable to widely-used tunable-coupler architectures in scalable processors. In these systems, direct spectroscopy of the coupler frequency is challenging due to the weak coupling strength between resonators and couplers or the need of extra calibration procedures~\cite{xtalk_coupler_sloshing_mode,coupler_readout_from_nearby_resonator,coupler_readout_dispersive,coupler_readout_swap}. Therefore, straightforward and high-precision flux crosstalk calibration of tunable couplers without direct coupler spectroscopy has remained elusive.

Here, we present and demonstrate two independent methods for characterizing flux crosstalk in tunable couplers without directly driving or measuring the couplers themselves. These methods target different stages of chip calibration. The first method requires no prior spectroscopy of either qubits or couplers, we name it zero-knowledge characterization (ZKC). It takes advantage of the existing coupling between couplers and nearby resonators. This method enables the crosstalk estimation prior to the coarse calibration stage, at which point qubit coherence control and spectral characterization have not yet been performed. It also facilitates the parallel characterization of chip at a later stage. The second method employs the symmetric branch cancellation~(SBC) to disentangle the crosstalk in the target coupler from other dispersive effects imposed by neighboring elements, enabling high precision characterization of flux crosstalk prior to the gate calibration stage. We further discuss an iterative calibration in the presence of residual crosstalk, and achieve the residual crosstalk of couplers below $5\times10^{-5}$ through an improved procedure. Combining the two methods with the iterative calibration procedure provides a complete solution for a rapid and accurate flux crosstalk calibration of tunable couplers from zero knowledge to high fidelity operations.

\begin{figure}[t]
    \includegraphics{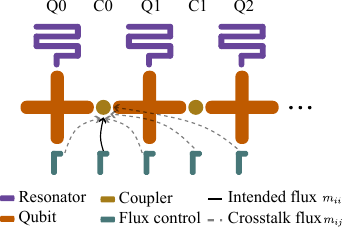}
    \caption{\textbf{The schematic of the device.} A subset of five elements are shown. The qubits and couplers form a linear chain, and each coupler is capacitively coupled to two neighbouring qubits. Dedicated readout resonators are capacitively coupled to the qubits. All elements are frequency tunable via individual flux control. However, due to flux crosstalk, each element receives both the intended flux (solid line) and the crosstalk flux~(dashed line).}
    \label{fig:fig1_main}
\end{figure}

The absence of dedicated readout channels makes characterization of couplers challenging. We circumvent this limitation by indirectly measuring the frequency shift of the couplers. This is achieved through existing coupling to nearby elements. We demonstrate our methods on a linear chain of ten transmon qubits and nine transmon couplers, and a subset of five elements is used for simplicity, as shown in Fig.~\ref{fig:fig1_main}. In this setup, the maximum coupler frequency is higher than the resonator frequency, which in turn exceeds the maximum qubit frequency. All qubits have their dedicated readout resonators that are capacitively coupled, whereas the couplers do not have readout channels. All qubits and couplers have SQUIDs with symmetrical junctions. The frequencies of qubits and couplers are flux tunable via individual flux control. 

Ideally, each tunable element should be under the sole control of its respective fluxline. However, the presence of the flux crosstalk compromises this independence. The tunable elements receive not only the intended flux from their dedicated fluxlines, but also crosstalk flux originating from other fluxlines. We refer to the element where the crosstalk flux originates as the ``probe", and the element that receives the crosstalk flux as the ``target". The behaviour of the flux crosstalk is quantified by the crosstalk matrix $M$, and the elements in the matrix are defined as 
\begin{equation}
    m_{ij} = \frac{\partial \phi_i}{\partial \phi_j},
\end{equation}
where $m_{ij}$ is the crosstalk coefficient from the probe $j$ to the target $i$, and $\phi_i$ is the flux received by element $i$.

In a tunable quantum system, it is desirable to calibrate the system in the absence of noticeable flux crosstalk even at the coarse calibration stage, as such crosstalk can induce unintended interactions among a large number of elements in the system. However, the direct spectroscopy of couplers without available  readout channels often requires a complicated calibration procedure~\cite{coupler_readout_swap,coupler_readout_dispersive}, which itself is susceptible to flux crosstalk. The ZKC method can suppress flux crosstalk even before any qubit spectroscopy is performed, and thus can act as the first step in the calibration procedure.

\begin{figure}[h]
    \centering
    \includegraphics{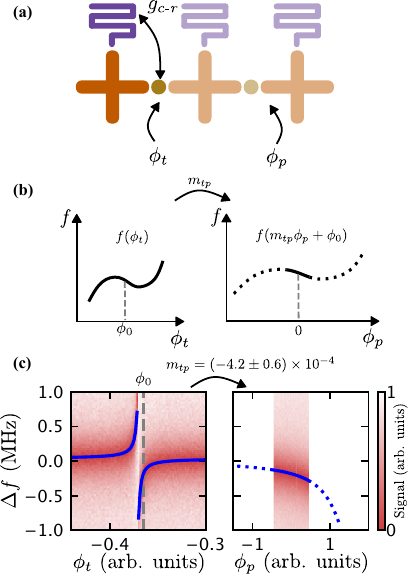}
    \caption{\textbf{Zero-knowledge characterization.} (a) The schematic of the device setup. The coupling $g_{c\text{-}r}$ is used for crosstalk calibration, with $\phi_t$ being the flux applied to target and $\phi_p$ being the flux applied to probe. The highlighted elements are involved in the calibration, including the auxiliary qubit with readout resonator and the target coupler. Any shaded elements can be chosen as the probe. (b) The schematics of a nonlinear frequency-flux dependence with target flux (left) and probe flux (right) under a crosstalk coefficient $m_{tp}$. The solid line indicates the measured frequency, and the dotted line represents the predicted frequency. $\phi_0$~indicates the idling flux for the target. (c) The resonator frequency-flux dependence with target flux (left) and probe flux (right). The color represents the resonator transmission, and the vertical grey dashed line indicates the chosen idling point. The crosstalk coefficient $m_{tp}=(-4.2\pm0.6)\times10^{-4}$ is fitted from the curves. }
    \label{fig:fig2_main}
\end{figure}

In practice, there is a small but useful coupling $g_{c\text{-}r}$ between the coupler and the readout resonators of nearby qubits, which we can utilize to characterize the flux crosstalk to the coupler, as shown in Fig.~\ref{fig:fig2_main}(a). In an architecture where the maximum coupler frequency is above the resonator frequency, such coupling will cause an antilevel crossing between the resonator and the coupler, and the induced frequency change in the resonator is governed by
\begin{equation}
    \Delta f(\phi_t) =\frac{\delta(\phi_t)}{2}\left(1-\sqrt{1+\left(\frac{2g_{c\text{-}r}}{\delta(\phi_t)}\right)^2}\right),
    \label{eq:antilevel}
\end{equation}
where $\phi_t$ is the flux at the coupler, and $\delta(\phi_t)=f_c(\phi_t)-f_r$ is the frequency difference between the resonator and the coupler. Due to the weak coupling strength and the offset flux of the coupler, the antilevel crossing feature is confined in a small flux region, and the frequency difference can be linearized by 
\begin{equation}
    \delta(\phi_t)\approx \frac{df_c}{d\phi_t}(\phi_t-\bar\phi),
\end{equation} 
with $\bar\phi$ being the coupler flux where $f_c(\bar\phi)=f_r$. We refer to the qubit that is strongly coupled to this resonator as the auxiliary qubit.

The ZKC method uses this antilevel crossing as the indicator for the crosstalk effect. The flux crosstalk is mathematically equivalent to a horizontal scaling and translation operation of the frequency-flux dependency as shown in the schematic Fig.~\ref{fig:fig2_main}(b). The crosstalk coefficient $m_{tp}$ corresponds to the scaling factor, and the idling flux $\phi_0$ of the target corresponds to the translation offset. 

In Fig.~\ref{fig:fig2_main}(c), we show an example of the crosstalk calibration with the ZKC method. We first sweep the target flux and measure the resonator response $\Delta f(\phi_t)$. By fitting $\Delta f(\phi_t)$ to Eq.~\ref{eq:antilevel}, we obtain the values of $g_{c\text{-}r}$, $f_r$, $\bar\phi$ and $df_c/d\phi_t$. The crosstalk curve is then measured by scanning the probe flux, and the crosstalk ratio $m_{tp}$ can be extracted by further fitting $\Delta f(m_{tp}\phi_p+\phi_0)$ to Eq.~\ref{eq:antilevel} with other parameters fixed. The solid lines represent the fitted frequency shift in the resonator, and the dashed lines represent the predicted frequency shift outside the measurable flux range. No direct drive or readout of either the qubit or the coupler is required during the experiment.

Applying the ZKC method to the couplers requires that it first be performed on the qubits (see Supplementary Note). This requirement arises because any frequency shift of the auxiliary qubit also shifts the resonator frequency, thereby degrading the accuracy of ZKC when applied to the couplers. Accordingly, we apply ZKC to the auxiliary qubit to calibrate the flux crosstalk to that qubit, and locate the flux point at which its frequency is minimized. With only two iterations, the residual crosstalk falls below 0.25\% for all but one matrix element (see Supplementary Note). In principle, the method is applicable across a broad range of architectures and coupling strengths, provided that there is a resolvable feature in the resonator-flux spectrum. Fig.~\ref{fig:fig2_main} demonstrates that crosstalk can be calibrated without any prior qubit or coupler spectroscopy, relying solely on the resonator–flux dependence, an advantage for the initial bring-up of large-scale processors.

After obtaining the initial crosstalk calibration result from the ZKC method, we now focus on finetuning individual crosstalk coefficients to high precision through the SBC method. This requires us to leverage the high precision measurement of the qubit frequency. The crosstalk to the coupler can induce a shift in the qubit frequency through the designed coupling $g_{c\text{-}q}$. At the same time, direct crosstalk to the qubit can also alter its frequency. Therefore, an accurate crosstalk calibration requires us to distinguish clearly between frequency shifts arising from the crosstalk to the qubit and those mediated through the crosstalk to the coupler. While the method previously proposed in Ref.~\cite{Symmetric} is effective for measuring flux crosstalk to qubits, it is not directly applicable to couplers. To overcome this limitation, we generalize the method to address a broader range of cases.  

\begin{figure}[t]
    \centering
    \includegraphics{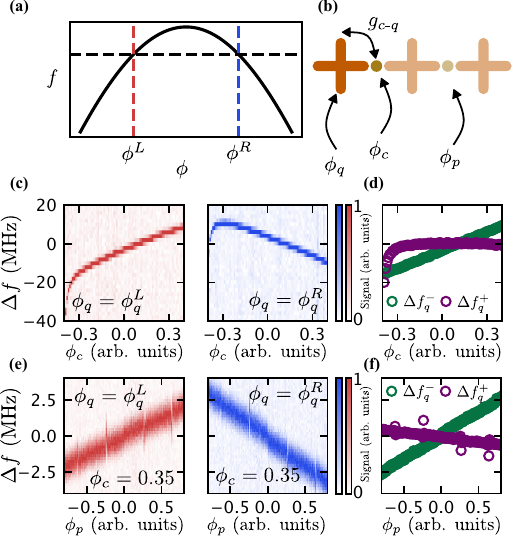}
    \caption{\textbf{Symmetric branch cancellation.} (a) The illustration of the symmetric transmon spectrum. $\phi^R$ and $\phi^L$ are the symmetric bias points where the frequencies are the same but the slopes have opposite signs. (b) The schematic of the device setup. The coupling $g_{c\text{-}q}$ is used for crosstalk calibration, with $\phi_q$ being the flux applied to the auxiliary qubit, $\phi_c$ being the flux applied to the target coupler, and $\phi_{p}$ being the flux applied to the probe. The highlighted elements are involved in the calibration, including the auxiliary qubit and the target coupler. Any shaded elements can be chosen as  the probe. (c) Qubit frequency as a function of $\phi_c$ when the qubit is bias to the left branch $\phi_q^L$ (left) and to the right branch $\phi_q^R$ (right). (d) The calculated $\Delta f_q^-(\phi_c)$ and $\Delta f_q^+(\phi_c)$ quantify the crosstalk and the dispersive effect from the coupler to the qubit, respectively. (e)(f) Same as in (c)(d) but as a function of $\phi_{p}$. The calculated $\Delta f_q^-(\phi_{p})$ and $\Delta f_q^+(\phi_{p})$ quantify the crosstalk from the probe to the qubit and to the coupler, respectively. }
    \label{fig:fig3_main}
\end{figure}

\begin{figure*}[!t]
    \includegraphics{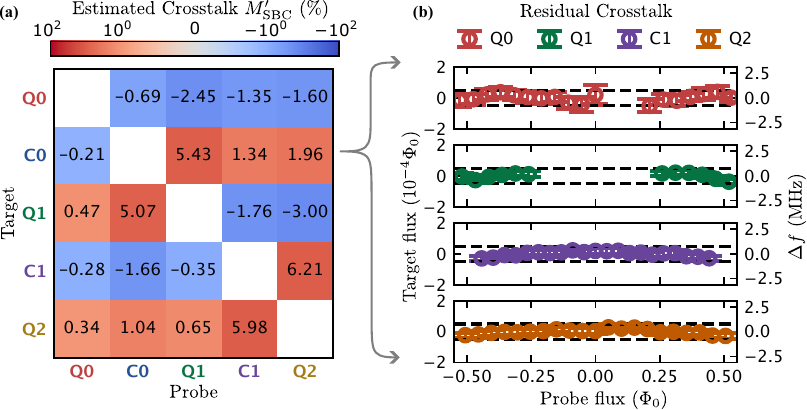}
    \caption{\textbf{Crosstalk characterization result.} (a) The estimated crosstalk matrix $M'_{\text{SBC}}$ after the SBC characterization method. (b) The measured residual crosstalk for C0 using other elements as the probe. The auxiliary qubit is Q1 when the probe is Q0, and the auxiliary qubit is Q0 when the probe is Q1, C1, or Q2. A linear fit to the data yields slopes below $5\times10^{-5}$. The dashed lines are $\pm5\times10^{-5}~\Phi_0$ received by the target.}
    \label{fig:fig4_main}
\end{figure*}
    
The key note behind the SBC is the symmetry of the transmon spectrum. If we assume that the residual crosstalk is sufficiently suppressed by prior ZKC calibration, the total frequency shift of a qubit that couples to a single coupler can be modeled and linearized around the idling point as 
\begin{align}
    \Delta f_q 
    &\approx\frac{\partial f_q}{\partial \phi_q}\Delta\phi_q + \frac{g_{c\text{-}q}^2}{\delta^2}\left(-\frac{\partial f_q}{\partial \phi_q}\Delta\phi_q+\frac{\partial f_c}{\partial \phi_c}\Delta\phi_c\right)\notag\\
    &\approx\frac{\partial f_q}{\partial \phi_q}\Delta\phi_q + \frac{g_{c\text{-}q}^2}{\delta^2}\frac{\partial f_c}{\partial \phi_c}\Delta\phi_c,
    \label{eq:SBC_base}
\end{align}
with the assumption $g_{c\text{-}q}^2/\delta^2\ll1$, where $g_{c\text{-}q}$ is the coupling strength between the qubit and the coupler, $\delta=f_q-f_c$ is the frequency difference on the idling point, and $\Delta\phi_q$ ($\Delta\phi_c$) is the received crosstalk flux at the qubit (coupler). We refer to this qubit as the auxiliary qubit. Due to the symmetric spectrum of the transmon, one can selectively flip the sign of ${\partial f_q}/{\partial \phi_q}$~(${\partial f_c}/{\partial \phi_c}$) by biasing the qubit (coupler) to the symmetric points $\phi^R$ and $\phi^L$, as shown in Fig.~\ref{fig:fig3_main}(a). If we measure the qubit frequency shift $\Delta f_q^R$ ($\Delta  f_q^L$) at bias $\phi_q^R$ ($\phi_q^L$) while sweeping the probe flux $\phi_{p}$~[Fig.~\ref{fig:fig3_main}(b)], we obtain 
\begin{align}
    \Delta f_q^-(\phi_{p})&=\frac{\Delta f_q^R-\Delta f_q^L}{2} = \frac{\partial f_q}{\partial \phi_q}m_{qp}\phi_{p},
    \label{eq:SBC_1}\\
    \Delta f_q^+(\phi_{p})&=\frac{\Delta f_q^R+\Delta f_q^L}{2}=\frac{g_{c\text{-}q}^2}{\delta^2}\frac{\partial f_c}{\partial \phi_c}m_{cp}\phi_{p},
    \label{eq:SBC_2}
\end{align}
where $\Delta f_q^-(\phi_{ex})$ and $\Delta f_q^+(\phi_{ex})$ quantify the crosstalk coefficients from the probe to the qubit and to the coupler, respectively. 

We show an example of SBC in Fig.~\ref{fig:fig3_main}(c)-(f), where the qubit frequency is measured using qubit spectroscopy. We first measure the frequency of the auxiliary qubit at the symmetric points while sweeping the coupler flux $\phi_c$, as shown in Fig.~\ref{fig:fig3_main}(c)-(d), which gives $\Delta f_q^-(\phi_c)$ that quantifies the $m_{qc}$, and $\Delta f_q^+(\phi_c)$ that can be utilized to calculate the sensitivity prefactor $g_{c\text{-}q}^2/\delta^2\cdot\partial f_c/\partial \phi_c$ in Eq.~\ref{eq:SBC_2} for the chosen idling point. Subsequently we repeat the same measurement protocol with the probe flux $\phi_{p}$ varied instead, as shown in Fig.~\ref{fig:fig3_main}(e)-(f). We obtain $\Delta f_q^-(\phi_{p})$ and $\Delta f_q^+(\phi_{p})$, which quantify the crosstalk to the qubit ($m_{qp}$) and the coupler ($m_{cp}$), respectively.

We can also apply the SBC method to couplers. In a case where the auxiliary qubit is connected to multiple couplers,
we need to flip the sign of ${\partial f_c}/{\partial \phi_c}$ in Eq.~\ref{eq:SBC_base} by instead biasing the target coupler to symmetric points $\phi_c^L$ and $\phi_c^R$. We therefore can selectively extract the crosstalk to the target coupler by obtaining 
    \begin{equation}
        \Delta f_q^-(\phi_{p})=\frac{\Delta f_q^R-\Delta f_q^L}{2} = \frac{g_{qc}^2}{\delta^2}\frac{\partial f_c}{\partial \phi_c}m_{cp}\phi_{p}.
    \end{equation}
Using the symmetric branch cancellation, Fig~\ref{fig:fig3_main} evidences that $\Delta f_q^+$ and $\Delta f_q^-$ separate out the frequency shifts induced by the crosstalk to the target and those arose from the crosstalk to other elements.

The crosstalk matrix characterization normally requires multiple iterations until the residual crosstalk is negligible. During this process, the flux will be compensated based on the estimated crosstalk matrix before it is applied to the chip. Let $M'$ be the estimated crosstalk matrix, and $\Phi$ be the flux we apply before the crosstalk compensation. The actual flux arriving on the chip is 
\begin{equation}
    \Phi'=M(M')^{-1}\Phi.
    \label{eq:compensation}
\end{equation}
For a single iteration, the crosstalk ${\partial\phi'_t}/{\partial\phi_p}$ will be measured, and the estimated crosstalk matrix will be updated. Using Eq.~\ref{eq:compensation}, we find that the measured residual crosstalk is a linear combination of the actual residual crosstalk from all elements to the target,
\begin{equation}
    \frac{\partial\phi'_t}{\partial\phi_p} = \sum_i\delta m_{ti}(M')_{ip}^{-1}.
    \label{eq:update}
\end{equation}
This indicates that the error of the estimated crosstalk matrix will propagate through the calibration process. The situation becomes more severe when a single element has a large error or when numerous small errors accumulate. It is preferable to first measure the residual crosstalk from all elements to the target, and then recover the actual residual crosstalk using Eq.~\ref{eq:update}. This can reduce the number of iterations needed for high precision crosstalk characterization, and consequently speedup the calibration process.

In Fig.~\ref{fig:fig4_main}, we demonstrate the flux crosstalk calibration results. We first employ the ZKC method for two iterations to calibrate the crosstalk coefficient, followed by two additional iterations using the SBC method to achieve high-precision crosstalk calibration. Then the residual crosstalk is measured by SBC method. The iterative calibration follows the update procedure described above. Fig.~\ref{fig:fig4_main}(a) shows the estimated crosstalk matrix $M'_{\text{SBC}}$ after the SBC method, and Fig.~\ref{fig:fig4_main}(b) shows the residual crosstalk for coupler C0 as a representative example (see Supplementary Note for other elements). The sensitivity prefactor $g_{c\text{-}q}^2/\delta^2\cdot\partial f_c/\partial \phi_c$ at the idling point is around $19~\text{GHz}/\Phi_0$, where the $\Phi_0$ is the magnetic flux quantum. A linear fit to the data yields slopes below $5\times10^{-5}$, which are the residual crosstalk coefficients. The dashed lines are $\pm5\times10^{-5}~\Phi_0$ flux received by the target, and it corresponds to a frequency shift of the auxiliary qubit within $\pm1$~MHz at this idling point, where the qubit is highly sensitive to flux.

In conclusion, we demonstrate two independent methods that efficiently characterize the flux crosstalk in a fully tunable architecture where the direct operation and measurement of couplers are not straightforward. Notably, the ZKC method allows an estimation of the crosstalk without any prior spectroscopy of either qubits or couplers, which benefits the initial bring-up for large scale processors, and facilitates the parallel characterization of chip at the later stage. In addition, the SBC method enables decoupling of flux crosstalk from different elements, and it is also applicable to high precision crosstalk characterization for couplers. Moreover, we improved the iterative calibration process, and achieve the residual crosstalk of couplers below $5\times10^{-5}$. These results offer a solution for flux crosstalk characterization of couplers from zero knowledge to high precision in a scalable and fully tunable architecture.
\\
\section*{ACKNOWLEDGMENTS}

The device is made in the Helmholtz Nano Facility. We acknowledge the support by the German Federal Ministry of Research, Technology and Space (BMFTR), funding program “Quantum technologies - from basic research to market”, project QSolid (Grant No. 13N16149). We also acknowledge the funding from the Deutsche Forschungsgemeinschaft (DFG) under Germany's Excellence Strategy-Cluster of Excellence Matter and Light for Quantum Computing (ML4Q) EXC 2004/2-390534769.

\section*{AUTHOR CONTRIBUTIONS}

Y.G., A.G. and R.B. designed the experiments. Y.G. performed the experiments and analyzed the data. Y.G., A.G. and J.C. implemented key components of the data-taking infrastructure. Y.G. designed the device. Y.L., Y.H., D.A.V., J.R.G., H.B., and M.N. contributed to device fabrication. Y.G., A.G., J.R.G., H.B., and M.J. contributed to the experimental infrastructures.  P.A.B. and R.B. supervised the project. Y.G. and R.B. prepared the manuscript. All authors contributed to the manuscript revision.

\section*{DATA AVAILABILITY}
The data that support the findings of this study are available from the corresponding author upon reasonable request.
\section*{COMPETING INTERESTS}
The authors declare no competing interests.

\end{document}


\title{Supplementary Materials for\\ ``Flux Crosstalk Calibration from Zero Knowledge to High Precision in Fully Tunable Superconducting Circuits''}

\author{Y. Gao }
\pgithirteen
\rwth
\author{A. Galicia}
\pgithirteen
\rwth
\author{J. R. Guimar\~aes}
\pgithirteen
\rwth
\author{Y.~Haddad}
\pgithirteen
\rwth
\author{Y. Liu }
\pgithirteen
\author{J. Cereijo}
\pgithirteen
\rwth
\author{D.~A.~Volkov}
\pgithirteen
\rwth
\author{H.~Bhardwaj}
\pgithirteen
\rwth
\author{M. Neis}
\pgithirteen
\rwth
\author{M. Jerger}
\pgithirteen
\author{P. A. Bushev}
\pgithirteen
\author{R. Barends}
\pgithirteen
\rwth

\maketitle

\renewcommand{\thefigure}{S\arabic{figure}}
\renewcommand{\thesection}{S\Roman{section}}
\renewcommand{\theequation}{S\arabic{equation}}
\section{Applying the ZKC method on qubits}

\begin{figure}[h]
    \includegraphics{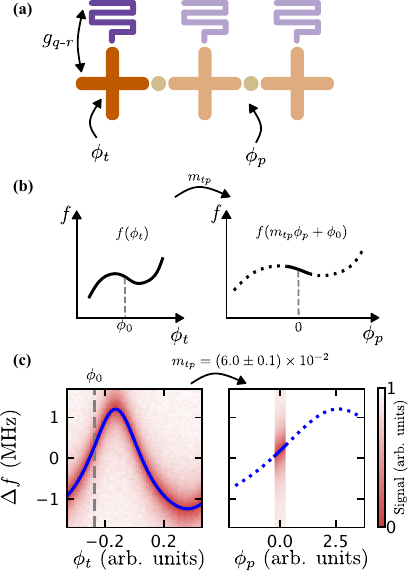}
    \caption{\textbf{ZKC method on qubits.} (a) The schematic of the device setup. The coupling $g_{q\text{-}r}$ is the qubit-resonator coupling strength, with $\phi_t$ being the flux applied to target and $\phi_p$ being the flux applied to probe. The highlighted elements are involved in the calibration, including the target qubit and the readout resonator. Any shaded elements can be chosen as the probe. (b) The schematics of a nonlinear frequency-flux dependence with target flux (left) and probe flux (right) under a crosstalk coefficient $m_{tp}$. The solid line indicates the measured frequency, and the dotted line represents the predicted frequency. $\phi_0$~indicates the idling flux for the target. (c) The measured resonator frequency-flux dependence with target flux (left) and probe flux (right). The color represents the resonator transmission, and the vertical grey dashed line indicates the chosen idling point. The crosstalk coefficient $m_{tp}=(6.0\pm0.1)\times10^{-2}$ is fitted from the curves. }
    \label{fig:zkc_qubit}
\end{figure}
Fig.~\ref{fig:zkc_qubit} demonstrates the application of ZKC method on qubits. It utilizes the resonator frequency shift caused by the qubit-resonator coupling $g_{q\text{-}r}$ as shown in Fig.~\ref{fig:zkc_qubit}(a). The resonator frequency shift is governed by \begin{equation}
    \Delta f(\phi_t) =\frac{\delta(\phi_t)}{2}\left(1-\sqrt{1+\left(\frac{2g_{q\text{-}r}}{\delta(\phi_t)}\right)^2}\right),
    \label{eq:antilevel}
\end{equation}
where $\phi_t$ is the flux at the qubit, and $\delta(\phi_t)=f_q(\phi_t)-f_r$ is the frequency difference between the resonator and the qubit. The qubit frequency $f_q(\phi_t)$ is modeled as 
\begin{equation}
    f_q(\phi_t) = f_q^{\text{max}}\sqrt{|\cos{(\pi\bar\phi_t)}|},
\end{equation}
where $\bar\phi_t$ is the normalized flux at the target qubit. Fig.~\ref{fig:zkc_qubit}(c) shows an example of crosstalk calibration performed on qubits with the ZKC method, following the procedure described in the main text. The crosstalk coefficient $m_{tp}=(6.0\pm0.1)\times10^{-2}$ for the target qubit is extracted from the fit. The flux point at which the qubit frequency is minimized is identified from the corresponding minimum in the resonator frequency, as the shift is minimum when the qubit-resonator detune is largest. 

\section{Residual crosstalk}
\begin{figure}[!b]
    \includegraphics{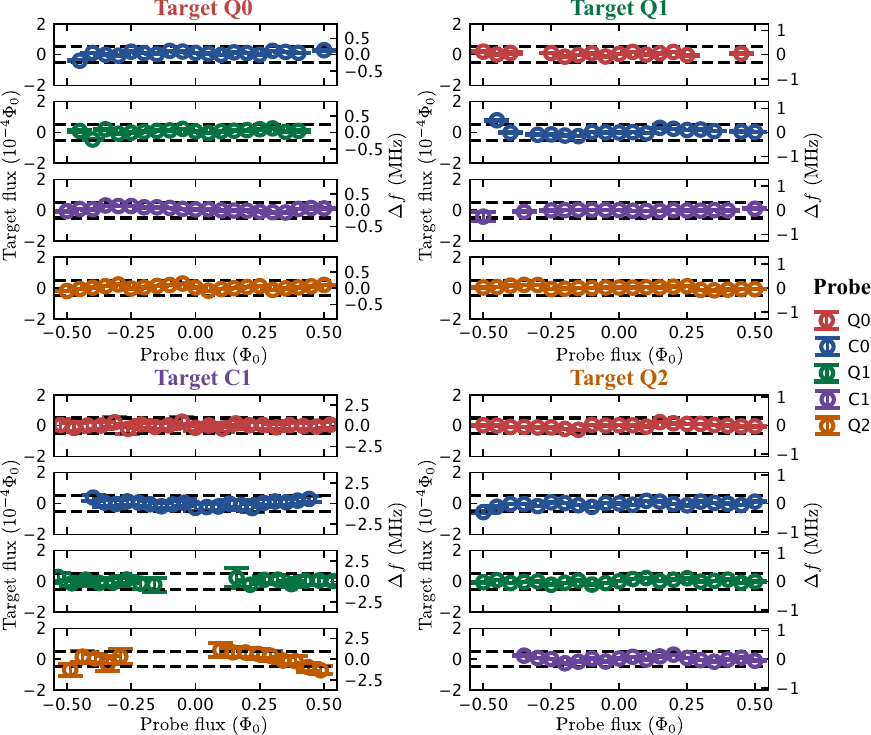}
    \caption{\textbf{Residual crosstalk.} The measured result crosstalk using the SBC method. Linear fits to the data yield all slopes below $5\times10^{-5}$, which corresponds to the crosstalk coefficients. The dashed lines are $\pm5\times10^{-5}~\Phi_0$ received by the target.}
    \label{fig:residual_summary}
\end{figure}

In Fig.~\ref{fig:residual_summary} we show the measured residual crosstalk of all elements using the SBC method. Linear fits to the data yield all slopes below $5\times10^{-5}$. The sensitivity prefactors $\partial f_q/\partial\phi_q$ are 4.5, 5.6, and 5.2~GHz/$\Phi_0$ for Q0, Q1, and Q2, respectively. The sensitivity prefactor $g_{c\text{-}q}^2/\delta^2\cdot\partial f_c/\partial \phi_c$ for C1 is around 15~GHz/$\Phi_0$. The dashed lines are $\pm5\times10^{-5}~\Phi_0$ flux received by the target. 

When the target is the coupler C1, the auxiliary qubit is Q1 if the probe is Q2, whereas the auxiliary qubit is Q2 if the probe is Q0, C0, or Q1. We use qubit spectroscopy to measure the auxiliary qubit frequency when the target is a coupler, and the idling point is chosen close to the qubit-coupler antilevel crossing, which enables the measurement to be performed with enhanced sensitivity. When the target is a qubit, we use Ramsey to measure the qubit frequency.

When the coupler is the target and a nearest-neighbour qubit is the probe, we exclude data whose standard error exceeds $5\times10^{-4}$, because the probe qubit perturbs the antilevel crossing feature which the calibration relies on, and moves the auxiliary qubit outside the frequency measurement window. This can be seen when C1 is the target in Fig.~\ref{fig:residual_summary} and when C0 is the target in Fig.~4 of the main text.

\section{Estimated crosstalk matrix from the ZKC method}

\begin{figure}[h]
    \includegraphics{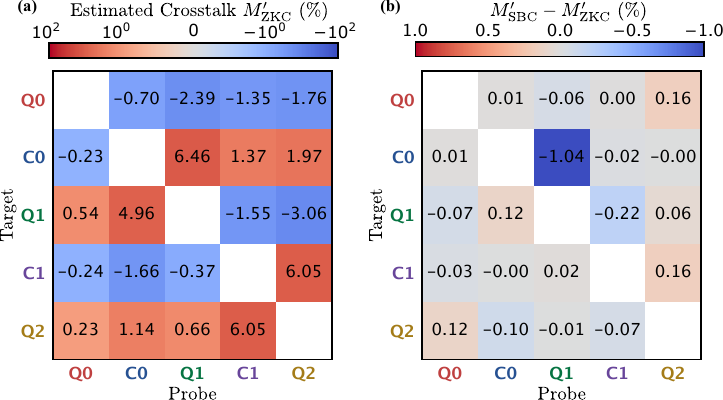}
    \caption{\textbf{Crosstalk characterization result after the ZKC method.} (a) The estimated crosstalk matrix $M'_{\text{ZKC}}$ after using the ZKC method for two iterations. (b) The difference between the $M'_{\text{SBC}}$ and  $M'_{\text{ZKC}}$. The residual crosstalk after the ZKC method is less than 0.25\% with only one exception.}
    \label{fig:ZKC_result}
\end{figure}

To further demonstrate the effectiveness of the ZKC method, we compare the estimated crosstalk matrix $M_{\text{ZKC}}'$ with the high-accuracy reference estimate $M_{\text{SBC}}'$ in Fig.~\ref{fig:ZKC_result}. The calibration procedure is initialized using the identity matrix and performed for two iterations using ZKC method. The resulting matrix $M_{\text{ZKC}}'$
is shown in Fig.\ref{fig:ZKC_result}(a), while the difference between 
$M_{\text{ZKC}}'$ and $M_{\text{SBC}}'$, the most accurate estimate obtained in this work, is plotted in Fig.\ref{fig:ZKC_result}(b). After two iterations of the ZKC method, the residual crosstalk remains below 0.25\% for all matrix elements except one.

\section{Optimal idling point with qubits being the target}

\begin{figure}[!h]
    \includegraphics{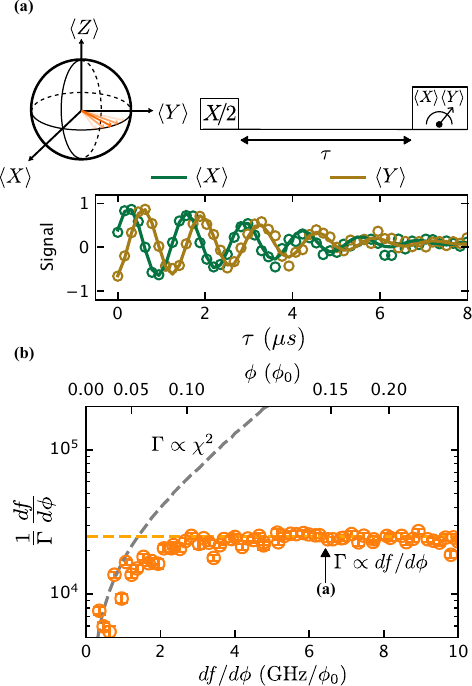}
    \caption{\textbf{The optimal idling point for qubits.} (a) The Ramsey experiment that is used for measuring the qubit frequency. (b) The minimum resolvable flux step at different idling points. The grey dashed line indicates the dephasing limitation from the resonator, and the orange dashed line represents the limitation arising from 1/f noise. The optimal idling point is achieved when the qubit dephasing rate is fully dominated by 1/f noise. The arrow indicates the data point plotted in (a).}
    \label{fig:optimal_point}
\end{figure}

High-accuracy crosstalk calibration requires resolving very small flux differences. In practice, this is achieved by measuring correspondingly small shifts in the qubit transition frequency using Ramsey experiment. If 
$\Delta f$ denotes the minimum resolvable frequency shift, the minimum resolvable flux step satisfies
\begin{equation}
    \Delta\phi^{-1}_{\text{min}} = \frac{1}{\Delta f}\frac{df}{d\phi},
\end{equation}
where ${df}/{d\phi}$ is the frequency-flux slope at the idling point. However, the frequency resolution is limited by the qubit dephasing rate $\Gamma$ in the Ramsey experiments, as shown in Fig.~\ref{fig:optimal_point}(a). Therefore the achievable flux sensitivity is bounded by  
\begin{equation}
    \Delta\phi^{-1}_{\text{min}} \propto \frac{1}{\Gamma}\frac{df}{d\phi}.
\end{equation}
Fig.~\ref{fig:optimal_point}(b) shows the inverse of the minimum resolvable flux step $\Delta\phi^{-1}_{\text{min}}$, evaluated at different idling points.  Larger values correspond to finer flux resolution. Two distinct regimes are apparent. When the qubit is close to the flux insensitive point, the dephasing rate is dominated by thermal photons in the readout resonator, and it scales as~\cite{photon_dephasing}
\begin{equation}
    \Gamma\propto\chi^2,
\end{equation} 
where $\chi$ is the dispersive shift in the resonator. When the qubit is away from the sweet spot, the dephasing rate is dominated primarily by $1/f$ noise, and it follows~\cite{f_dephasing}
\begin{equation}
    \Gamma\propto\frac{df}{d\phi}.
\end{equation} 
The optimal idling point for measuring qubit flux crosstalk can be achieved when the qubit dephasing rate is fully dominated by 1/f noise.

We note that the above analysis does not apply when the target element is a coupler, because the assumptions underlying the dephasing-limited estimate break down in the vicinity of a qubit-coupler anticrossing. Experimentally, we instead find that qubit spectroscopy provides a smaller minimum resolvable flux step when the target element is a coupler.